\documentclass[aip,sd,amsmath,amssymb,
 reprint,%
]{revtex4-1}

\usepackage{dcolumn}% Align table columns on decimal point
\usepackage{bm}
\usepackage{graphics}
\usepackage{graphicx}
\usepackage{epsfig}
\usepackage{hyperref}
\usepackage{color}
\usepackage{amsfonts}
\usepackage[normalem]{ulem}
\usepackage{float}
\newcommand{\beq}{\begin{equation}}
\newcommand{\eeq}{\end{equation}}
\newcommand{\beqar}{\begin{eqnarray}}
\newcommand{\eeqar}{\end{eqnarray}}
\newcommand{\bcen}{\begin{center}}
\newcommand{\ecen}{\end{center}}

\providecommand\sca[2]{\langle #1 | #2 \rangle}

\begin{document}

\title[Stochastic stimulated electronic x-ray Raman spectroscopy]{Stochastic stimulated electronic x-ray Raman spectroscopy}
%Coherent stimulated x-ray Raman spectroscopy driven by stochastic free-electron laser radiation}
\author{Victor Kimberg}\email{Kimberg@kth.se}
\affiliation{Theoretical Chemistry and Biology, Royal Institute of Technology, 10691 Stockholm, Sweden}
\affiliation{Max Planck Institute for the Physics of Complex Systems, Noethnitzer Str. 38, 01187 Dresden, Germany}
\author{Nina Rohringer}\email{Nina.Rohringer@mpsd.mpg.de}
\affiliation{Max Planck Institute for the Physics of Complex Systems, Noethnitzer Str. 38, 01187 Dresden, Germany}
\affiliation{Center for Free-Electron Laser Science, Luruper Chaussee 149, 22761 Hamburg, Germany}
\affiliation{Max Planck Institute for the Structure and Dynamics of Matter, Luruper Chaussee 149, 22761 Hamburg, Germany}

\begin{abstract} 
Resonant inelastic x-ray scattering (RIXS) is a well-established tool for studying electronic, nuclear and collective dynamics of excited atoms, molecules and solids. An extension of this powerful method to a time-resolved probe technique at x-ray free electron lasers (XFELs) to ultimately unravel ultrafast chemical and structural changes on a femtosecond time scale is often challenging, due to the small signal rate in conventional implementations at XFELs that rely on the usage of a monochromator set up to select a small frequency band of the broadband, spectrally incoherent XFEL radiation. Here, we suggest an alternative approach, based on stochastic spectroscopy, that uses the full bandwidth of the incoming XFEL pulses. Our proposed method is relying on stimulated resonant inelastic x-ray scattering, where in addition to a pump pulse that resonantly excites the system a probe pulse on a specific electronic inelastic transition is provided, that serves as seed in the stimulated scattering process. The limited spectral coherence of the XFEL radiation defines the energy resolution in this process and stimulated RIXS spectra of high resolution can be obtained by covariance analysis of the transmitted spectra.  We present a detailed feasibility study and predict signal strengths for realistic XFEL parameters for the CO molecule resonantly pumped at the O$1s\rightarrow\pi^*$ transition. Our theoretical model describes the evolution of the spectral and temporal characteristics of the transmitted x-ray radiation, by solving the equation of motion for the electronic and vibrational degrees of freedom of the system self consistently with the propagation by Maxwell’s equations.  
\end{abstract}

\date{\today}

%\pacs{xxx}% PACS, the Physics and Astronomy
%\keywords{Stimulated x-ray emission, x-ray free-electron laser, resonant inelastic x-ray spectroscopy, covariance analysis} %Use showkeys class option if keyword
\maketitle

\section{Introduction}

X-ray free-electron lasers (XFELs), delivering high-brilliance x-ray pulses of femtosecond (fs) duration, have the potential to revolutionize our ways to probe chemical reaction dynamics and follow structural changes on the spatial and temporal scale of nuclear and electron motion. These structural changes, mostly induced by coherent pump sources in the optical, IR, THz, but also x-ray spectral region, are typically followed by analyzing the change of the electron density, as for example in time-resolved implementations of x-ray diffraction, or femtosecond serial crystallography \cite{Tenboer05122014,Barends10092015}. Time-resolved x-ray emission \cite{zhang_gaffney} or absorption spectroscopy \cite{bressler1,doi:10.1146/annurev.physchem.56.092503.141310}, time-resolved resonant inelastic x-ray scattering \cite{Beye2013172,wernet2015} and time-resolved photoelectron \cite{perfetti2007ultrafast, 1367-2630-10-3-033004} or Auger spectroscopy \cite{guehr1,zhangrohringer} are complementary analysis tools to study chemical and structural changes. The combination of x-ray emission spectroscopy with x-ray crystallography within one experimental setup allows the study of both structural and chemical dynamics \cite{kern1, Kern201587} at the same experimental conditions and might evolve into a powerful tool. Time-resolved x-ray emission spectroscopy of chemically and biologically relevant samples, however, often suffer of small signal rates, that could be overcome by coherent amplification of the signal by stimulated x-ray emission \cite{rohringer_atomic_2012, kimberg_amplified_2013, yoneda}. Most notably, pump-probe experiments could take advantage of recording good-quality spectra under a single XFEL exposure. Here, we address the question of coherent signal amplification of a spectroscopic, photon-in photon-out technique -- resonant inelastic x-ray scattering (RIXS) -- with XFEL sources and present a feasibility study on stimulated inelastic x-ray scattering of molecular gases with intrinsically incoherent self-amplified spontaneous emission (SASE) radiation of XFELs. Our studies aim for quantitative predictive signal estimates, to critically assess new opportunities of stimulated inelastic x-ray scattering at present-day XFEL sources.

RIXS is a widely applied spectroscopic technique, that probes both un-occupied and occupied electronic states, and is sensitive to electronic, vibrational and elementary collective excitations \cite{RevModPhys.83.705}. Applications are wide, ranging from solid state physics \cite{RevModPhys.73.203,RevModPhys.82.847}, to high-resolution vibrational spectroscopy of molecules in the gas phase \cite{PhysRevLett.104.193002,PhysRevLett.106.153004}, hydrogen bonding in liquids \cite{PhysRevLett.114.088302}, and studies of charge transfer \cite{RevModPhys.74.703, B719546J}, etc. Compared to inelastic x-ray scattering far above any absorption edges, RIXS adds the advantage of element specificity of the scattering process, by tuning the photon energy of the incoming beam close to inner-shell ionization edges of a specific element in a molecular complex. In that way, catalytic reaction centers in large molecular complexes, often containing heavy elements and metals, can be specifically probed, and relevant chemical information, for example the oxidation state of a particular element in a catalytic reaction, or changes of the valence in charge-transfer processes, can be measured by a sensitive method.

With the invention of XFELs, providing x-ray pulses of fs duration, inelastic x-ray scattering is therefore becoming a viable tool to study the structural dynamics of optically induced changes in the gas, liquid and solid phase and at interfaces with fs time resolution. The information wealth of RIXS lies in the scanning of the incoming photon energy in fine steps through absorption resonances and absorption bands and recording of the emission spectrum with high energy resolution \cite{gelmukhanov_resonant_1999}. Being a photon-in photon-out technique, the information is typically presented in a two-dimensional map, which displays
a full RIXS spectrum for series of excitation spectrum in a conventional x-ray absorption spectroscopy scan. At third-generation x-ray sources based experiments, RIXS measurements are performed by using a high-resolution monochromator. At XFEL sources, the use of monochromators requires averageing over many x-ray pulses, since the effective number of photons in the monochromator band pass is highly fluctuating. A combination of self-seeded XFEL pulses \cite{PhysRevLett.114.054801,amann} and monochromators is a practicable experimental setting to overcome this deficiency for high-resolution RIXS studies at XFEL \cite{wernet2015}. Here, we want to address stimulated RIXS (SRIXS), i.e. stimulated resonant inelastic (Raman) x-ray scattering, as an alternative route, to record high-resolution RIXS spectra.

Stimulated electronic x-ray Raman scattering was recently realized in atomic neon gas \cite{Weninger2013e}, demonstrating coherent signal amplification by several orders of magnitude (as compared to spontaneous RIXS). Due to the stochastic nature of SASE pulses, a refined theoretical treatment \cite{Weninger2013d} was necessary, to unambiguously demonstrate this effect. In usual RIXS measurements, the emission wavelength shows a linear dispersion with respect to the incoming photon energy of a narrow-band x-ray source. A typical RIXS spectrum with a SASE spectral profile, consisting of mutually phase-uncorrelated spectral intensity spikes, can be understood as an incoherent sum of RIXS spectra of narrow-band sources with different detuning from the studied resonance. A RIXS spectrum recorded with a SASE pulse therefore shows strong shot-to shot fluctuations, with line shifts that are of stochastic nature. A covariance analysis of an ensemble of single-shot spectra, however, can be applied to reveal 2D maps similar to high-resolution RIXS maps.  The goal of this work is to extend SRIXS to molecular gas targets.

\begin{figure}
\centering
  \includegraphics[width =0.45\textwidth]{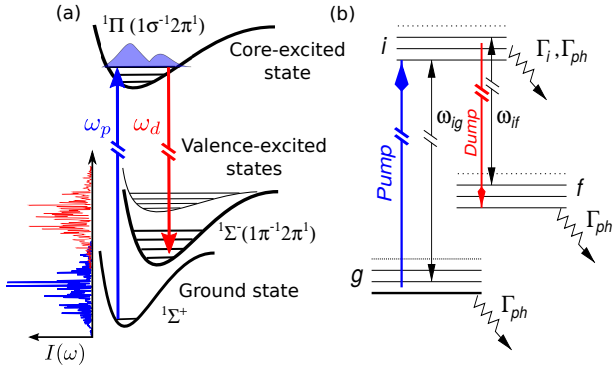}
  \caption{Two-color stimulated RIXS scheme driven by a pair of temporally overlapping SASE pulses. a) The pump field of frequency $\omega_{p}$ is resonant to the intermediate core-excited state $^1\Pi$ of CO; the probe field of frequency $\omega_{d}$ stimulates inelastic Raman scattering channel from the core-excited to a valence-excited state $^1\Sigma^-$. b) Generalized electronic-vibrational level scheme of stimulated RIXS in molecules.}
  \label{fig1}
\end{figure}

Similar to the x-ray lasing (amplified spontaneous emission) in molecular targets \cite{kimberg_amplified_2013}, that typically shows considerably lower gain as compared to the atomic case, Raman gain, or SRIXS cross sections in molecules tend to be considerably lower. Reasons for the smaller gain are the addition of vibrational and rotational degrees of freedom in molecular samples: The electronic transition dipole matrix elements in molecules are lowered by Franck-Condon overlaps. The gain is distributed over many vibrational channels. In an unaligned ensemble of molecules, there is a mismatch of the molecular transition dipoles with the emitted radiation, resulting in smaller stimulated emission cross sections. A way to circumvent the problem of small SRIXS cross section in molecular targets, is to provide seed photons to stimulated the scattering process. In our previous SRIXS experiment in atomic neon \cite{Weninger2013e} seed photons were provided by the spectral tails of the relatively broad SASE pulses. An alternative is to operate the FEL in recently established two-color modes \cite{Lutman2013,amann}, providing two narrow-band SASE pulses within the usual broadband SASE gain bandwidth (up to $\delta\omega/ \omega \approx$ 0.1-1 \%). The two central SASE energies should therefore overlap with the energy of the pump resonance and the inelastic electronic transition (“dump” transition) one wants to amplify (see Fig. \ref{fig1}). Since a core-excited state can typically decay to different electronic final states, by varying the dump transition energy, different valence-excited electronic states can be probed selectively. This is of special interest for “dark” emission channels, that in conventional RIXS experiments are hidden by strong neighboring transition lines. A single pair of SASE pulses will certainly couple to many different vibrational levels within a single pulse. A much clearer situation could be achieved by applying two narrow-band transform limited pulses. Two-color seeded FEL schemes are, however, currently available only at the EUV FEL source FERMI in Trieste \cite{allaria}. 

Here we discuss a statistical approach, based on a covariance analysis of the acquired SRIXS spectra in the two-color SASE scheme, that enables to unravel the rich vibrational structure of the involved electronic states. The high-energy resolution on the vibrational level using broad-band SASE pulses is possible due to the limited spectral coherence (average spike width in the spectral domain of the structured SASE pulses) \cite{Weninger2013d}. Statistically speaking, the covariance method provides SRIXS spectra that would have been obtained by varying pairs of coherent, narrow-band sources, with a spectral width that is smaller than the vibrational energy spacing and the core-hole lifetime. The SASE pulse-parameter regime, hence, does not allow coherent impulsive stimulated Raman scattering, to pump coherent vibrational or electronic wave packets, as discussed in applications of impulsive SRIXS (or as stimulated electronic –ray Raman scattering) in the many different pulse-schemes for nonlinear x-ray spectroscopy \cite{PhysRevLett.89.043001,doi:10.1021/jz501966h}.

The paper is organized as follows: in Sec. \ref{sec:general} we present the two-color scheme of SRIXS with SASE x-ray pulses; in Sec. \ref{sec:model} the theoretical model and the numerical approach are outlined; Results on the CO molecular are discussed in in Sec. \ref{sec:res}, and the method of covariance analysis is introduced (Sec. \ref{sec:low}); a critical feasibility study at present-day XFEL sources is presented in Sec. \ref{sec:exp}.

\section{Stimulated RIXS schemes driven by XFEL radiation}\label{sec:general}

High-resolution SRIXS spectra can be achieved in principle using a two-color x-ray source of well defined discrete frequencies $\omega_p$, tuned to the energy of a specific core-excited electronic state, and $\omega_d$, which probes inelastic scattering channels to a specific valence excited final state. Independent variation of $\omega_p$ and $\omega_d$ allows to build a two-dimensional SRIXS map $\sigma(\omega_p,\omega_d)$, similar to the usual RIXS maps recorded at 3\textsuperscript{rd} generation x-ray light sources. In the case of weak fields the information content that can be gained in this scheme is comparable to a conventional, spontaneous RIXS measurement at a synchrotron source, if the initial state of the system is the ground state.  In conventional RIXS the full emission spectrum can be collected for one fixed incoming photon energy $\omega_p$, while the SRIXS method with a coherent x-ray source of two well-defined frequencies $\omega_p$ and $\omega_d$ requires additional scanning of the energy $\omega_d$ and hence does not provide any advantages compared to conventional time-resolved RIXS measurements at XFELs \cite{wernet2015}.

Here we apply the two-color SASE XFEL scheme for a benchmark study of carbon-monoxide (CO), representing the class of abundant $\pi-$conjugated systems \cite{Sham1989, Skytt1997a}. Near-edge x-ray absorption fine structure (NEXAFS) spectra at the K-edge of many oxygen, nitrogen and carbon containing molecules conjugated by p-orbitals show a strong pre-edge resonance related to excitation of the $1s$ electron to the lowest unoccupied molecular orbital \cite{Sham1989}. This bound-to-bound transition dominates over other pre-edge resonances and coupling to the electronic continuum. The transition is typically isolated from the Rydberg resonances, thus ensuring x-ray scattering via an isolated intermediate electronic state, and making it particularly interesting for the study of valence-excited states of neutral molecules by RIXS. In the present paper we focus on the numerical simulations for the OK-edge of CO. The developed method, however, are general and can be directly applied to the CK-edge of CO, NK- and OK-edges of N$_2$, O$_2$, NO and other diatomic and small polyatomic molecules.

In conventional OK pre-edge RIXS spectra in CO, three partially overlapping excited valence states  -- $^1\Pi$, $^1\Delta$, and $^1\Sigma^{-}$  -- have been identified, by resonant scattering via the $1s$-$\pi^*$ resonance  \cite{Skytt1997a}. Stimulated x-ray scattering with a pair of  narrow-band SASE pulses allows to particularly address one of these valence excited electronic final states. In an unaligned ensemble of CO molecules in the ground state, the linear polarized XFEL pump pulse tuned to the  $1s$-$\pi^*$ resonance preferentially couples to a sub-ensemble of molecules with their axis aligned orthogonal to the E-field polarization vector ( the $\Sigma-\Pi$ transition dipole moment is orthogonal to the molecular axis). The dump pulse, having the same polarization direction as the pump, stimulates emission from the core-excited sub-ensemble of CO$^*$ molecules mainly on the $\Pi^*\rightarrow \Sigma, \Delta$ transitions, while the SRIXS cross section to the $\Pi$ final state is suppressed due to the dipole interaction symmetry. The potentials of $^1\Sigma^-$ and $^1\Delta$ states are almost parallel, and therefore have a very similar vibrational RIXS profile. In our current analysis, we restrict our study to a single final state, the $^1\Sigma^{-}$, that will dominate the SRIXS spectrum. 

\section{Theoretical model}\label{sec:model}

In order to get realistic signal estimates and understand propagation effects of the SRIXS process, we employ a theoretical scheme that follows the evolution of the molecular system, including vibrational and electronic degrees of freedom self consistently with the evolution of the electric field as a function of time and propagation distance through the medium. The theoretical scheme, is a generalization of previous work \cite{PhysRevA.90.063828,Weninger2013d,kimberg_amplified_2013}. The optically pumped medium can be considered as one-dimensional, given the geometry of the typical setup \cite{rohringer_atomic_2012,Weninger2013e}, that is roughly of cylindrical shape of large aspect ratio: In a typical experimental setup pulses are focused to a spot size of a few $\mu$m, while the focal depth, limiting the length of the gain medium, confined in a gas cell is typically a few mm. We suppose transverse homogeneity of the focused XFEL beam and therefore can use the paraxial wave approximation for modeling the x-ray field. Due to the relatively large energy separation of the two SASE components, we split the total electric field in two parts $\widetilde E(z,t)= \widetilde E_p(z,t)+\widetilde E_d(z,t)$, describing the strong field component on the pump transition $\widetilde E_p(z,t)$ and the dump field $\widetilde E_d(z,t)$. The $z$-direction is defined as the propagation direction of the XFEL pulse, and the field is supposed to be linearly polarized along the $x$-direction. The electric field components $\widetilde{E}_j(z,t)$ ($j=d,p$) are treated in the slow-varying envelope approximation and are expanded in terms of their driving frequencies $\omega_j$: 
\begin{equation}\label{etot}
\widetilde E_j(t,z)=E_j(t,z) e^{i(\omega_j t -k_j z)} + E_j^*(t,z) e^{-i(\omega_j t -k_j z)}
\end{equation}
For the driving frequencies we chose $\omega_p=534$ eV and $\omega_d=525$ eV. The SASE radiation is modeled as Gaussian noise \cite{Saldin1998383, PhysRevSTAB.6.050701} of a Gaussian averaged spectrum of a specific width (here, we suppose a spectral width of 5 eV at full width half maximum of the intensity). A numerical implementation to generate stochastic slowly varying field-envelopes $E_j(z=0,t)$ for the input fields can be found in References  \cite{Vannucci1980, PhysRevA.76.033416}. Within the paraxial approximation, the slow varying, complex envelopes $E_j(t,z)$ are evolving by \cite{Fleck1970}:
\begin{equation}\label{weq}
\frac{\partial E_j }{\partial z}+\frac{1}{c}\frac{\partial E_j}{\partial t}=
-\imath\frac{ 2\pi N}{c\omega_j} {\cal P}_j - \frac{\sigma_{ph} N}{2} E_j,
\end{equation}
where $c$ is the speed of light in vacuum and $N$ is the molecular density. The second term of the right hand side of Eq. (\ref{weq}) describes non-resonant absorption of the x-ray field due to photoionization of the valence electrons and the Carbon 1s electrons, governed by the total photoionization cross section $\sigma_{ph}(\omega_j)$. The first term of the right hand side of Eq.~(\ref{weq}) describes the resonant coupling of the field to the different electronic/vibrational transitions through the macroscopic polarization ${\cal P}_j$. The macroscopic polarization is connected to the density matrix of the molecular system in the usual way. The slow varying component ${\cal P}_j$ of the polarization in the rotating wave approximation (near resonance with the field frequency $\omega_j$) is given by
\begin{eqnarray}\label{pol1}
{\cal P}_j=\rho_{gi}d_{ig}\omega_{ig} e^{\imath(\omega_{ig}-\omega_{j})t}
+\rho_{fi}d_{if}\omega_{if} e^{\imath(\omega_{if}-\omega_{j})t}.
\end{eqnarray}
The molecular system is described by the density matrix in slowly varying amplitude approximation $\tilde\rho_{jk}\equiv \rho_{jk}e^{\imath\omega_{kj}t}$, where  $j=\{j_e,\nu_j\}$ denotes a composed index referencing the electronic state $j_e$ and vibrational quantum number $\nu_j$ (the index $k$ is defined accordingly). We explicitly treat three electronic states, the electronic ground state denoted by $g$, the intermediate O1s core-excited state $i$ and the final electronic state $f$. Since the applied pump and dump fields and the transitions $g\rightarrow i$ and $i\rightarrow f$ are well separated in frequency, we suppose that the pump field $E_p$ only resonantly couples the vibrational manifolds of $g$ and $i$ and the dump field $E_d$ only directly couples $i$ and $f$ (see Fig. \ref{fig1}(b)). In that way, the equation of motion of the density matrix (Liouville -- von Neumann equation) can be simplified by applying the rotating wave approximation and the numerical solution of the coupled differential equations becomes efficient and converges fast. The off-diagonal matrix elements of the slowly varying envelope of the density matrix are governed by the following equations (in atomic units):
\begin{eqnarray}\label{dm1}
\hat L_{ig}\rho_{ig} =  \sum_{g'}R_{ig'}^*\rho_{g'g}-\sum_{i'}\rho_{ii'}R_{i'g}^*+\sum_{f}R_{if}^*\rho_{fg},\nonumber\\
\hat L_{if}\rho_{if} =  \sum_{f'}R_{if'}^*\rho_{f'f}-\sum_{i'}\rho_{ii'}R_{i'f}^*+\sum_{g}R_{ig}^*\rho_{gf},\nonumber\\
\hat L_{fg}\rho_{fg} =  \sum_{i}R_{if}\rho_{ig}-\sum_{i}\rho_{fi}R_{ig}^*,\\
\hat L_{gg}\rho_{gg'} =  \sum_{i}R_{ig}\rho_{ig'}-\sum_{i}\rho_{gi}R_{ig'}^*,\nonumber\\
\hat L_{ff}\rho_{ff'} =  \sum_{i}R_{if}\rho_{if'}-\sum_{i}\rho_{fi}R_{if'}^*,\nonumber\\
\hat L_{ii}\rho_{ii'} = \sum_{f}R_{if}^*\rho_{fi'}-\sum_{f}\rho_{if}R_{fi'}^*\nonumber\\+\sum_{g}R_{ig}^*\rho_{gi'}- \sum_{g}\rho_{ig}R_{i'g},\nonumber\;,
\end{eqnarray}
where we defined
\begin{eqnarray}\label{l1}
\hat L_{lk}=-\imath\left(\frac{d}{dt}+\Gamma_{lk}\right ),\; l,k=g,i,f,
\end{eqnarray}
with the decay rates $\Gamma_{lk}$ defined as
\begin{eqnarray}
\Gamma_{ll}=\Gamma_l+\Gamma_{ph}^l,\; l=g,i,f, \\
\Gamma_{lk}=(\Gamma_l+\Gamma_k+\Gamma_{ph}^l+\Gamma_{ph}^k)/2,\; l,k=g,i,f,\nonumber\;.
\end{eqnarray}
We defined the complex Rabi frequencies $R_{ig}$ and $R_{if}$
\begin{eqnarray}\label{l2}
R_{ig}(t,z)={E_p(t,z)}{d}_{ig}e^{\imath(\omega_p-\omega_{ig})t},\nonumber\\
R_{if}(t,z)={E_d(t,z)}{d}_{if}e^{\imath(\omega_d-\omega_{if})t},
\end{eqnarray}
and $\omega_{ig}$ ($\omega_{if}$) is the transition frequency between vibrational sublevels of the ground $g$ (final $f$) and the intermediate $i$ states (see Fig.\ref{fig1}(b)). The transition dipole moment $d_{ij}= {{d}_{{ij},e}} \sca{\nu_i}{\nu_f}\; (j=g,f)$ is a product of the total electronic transition dipole moment and the vibrational Franck-Condon factors. For the CO molecule, the purely electronic transition dipole moments are fixed by by $d_{gi,e}=0.05$ a.u., $d_{if,e}=0.03$ a.u., as determined by an {\it ab initio} CASSCF calculation using the MOLPRO package \cite{molpro}. The Franck-Condon factor analysis is performed with the help of the solution of the stationary Schr\"odinger equation using the potential energy curves from Ref. \cite{Skytt1997a}. Let us underline, that the dipole interaction also depends on the angle $\theta$ between the electric field polarization and the transition dipole moment of the molecule $({\bf E\cdot d}_{ij})=E d_{ij}\cos(\theta)$. In order to compute the polarization Eq. (\ref{pol1}) of the system one actually has to average over a randomly oriented molecular ensemble \cite{kimberg_x-ray_2013}. This approach is numerically expensive, since the Eqs.(\ref{dm1}) have to be solved independently for numerous values of the angle $\theta$. In order to reduce computational costs, in the present model we use a reasonable approximation replacing the transition dipole moments by their rotational averages $d_{ij,e}\equiv \langle {d}_{ij,e}\rangle_\zeta$, where $\langle ... \rangle_\zeta$ means average over angle $\zeta$ between the molecular axis and polarization vector of the electric field. For the $\Sigma-\Pi$ transitions discussed here $\langle {d}_{ij,e}\rangle_\zeta^2={d}_{ij,e}^2 \langle\sin^2\zeta\rangle={d}_{ij}^2 2/3$.

We treat the decay of the density matrix elements by photo ionization, introducing a time and propagation-distance dependent decay rate
\begin{equation}
\Gamma_{ph}=\sigma_{ph}\Phi(z,t),
\end{equation}
where $\sigma_{ph}$ is approximated by the total ionization cross section at the photon energy of the pump pulse including ionization of the C 1s and valence electrons, and $\Phi(z,t)$ is the flux of the x-ray field. Specifically, we use $\sigma_{ph}=0.2$ Mb, which was estimated using atomic values for the cross-sections $\sigma_{ph}(C1s)\approx 0.17$ Mb, $\sigma_{ph}(2p,2s)\approx 0.03$ Mb \cite{Yeh1985}. The Auger decay of the intermediate core-excited state in Eqs. (\ref{dm1}) is treated phenomenologically by the decay rate $\Gamma_i=0.16$ eV (FWHM) \cite{Neeb1994c}. Since we neglect electron collisions at the considered rather low gas densities, the decay rate of the ground state (other than photoionization) is $\Gamma_{g}=0$. The radiative decay of the valence-excited final states, that are dipole forbidden, are neglected ($\Gamma_{f}=0$), since the decay times are long compared to all other time scales, and decay by photoionization ($\Gamma_{ph}$) is dominating.

Although the overall decrease in the spectral intensity by above threshold absorption (absorption involving transitions from the valence electrons and Carbon 1s shell) is considered in our model, the model does not include resonance absorption of O 1s electrons in molecular ions. The resonant Auger decay and thus the production of molecular ions is still the dominant decay process in the system, despite stimulated inelastic scattering. In our recent experiment \cite{lcls14expTBP} absorption dips of molecular ions were observed in the transmitted spectrum. However, only a few resonant transitions of molecular CO ions are known in the literature \cite{Skytt1997a} and {\it ab initio} calculations of the energy position and strength of the O 1s core-excited resonance position in molecular ions are not accurate enough. This makes it thus very challenging to identify the resonances that were observed in absorption and to quantitatively predict the changes of the transmitted spectrum by the diversity of absorption lines. The resonant absorption of the radiation by molecular and atomic ions is therefore omitted in our model. The experimental conditions (mostly the optical depth) have to be optimized to maximize the Raman gain while minimizing spectral features resulting of the resonance absorption. 

%\subsection{Details of the numerical simulations}
We solved equations (\ref{weq})-(\ref{dm1}) numerically using a finite-difference scheme for the wave equation and the 4$^{th}$ order Adams method \cite{press} for the density matrix equations. That implementation optimizes the computation time, since the convergence for the off resonant component requires a small integration time step. In the present case the results are converged with a time step of $\sim 10^{-2}$ fs.

In this study, we fixed the molecular density to  $N=2.5\times 10^{-19}$ cm$^{-3}$, corresponding to the typical experimental conditions. With a flat, cylindrical focus profile, the 1D model system is invariant with respect to rescaling of the density and length of the system, under the constraint of keeping the optical density (density-length product) constant. In order to reduce CPU time, we rescaled the density and length of the medium as compared to the experimental conditions:  the molecular density  $N'=N\times 10^3$  was increased by a factor of 1000, whereas the length of the medium was downscaled by a factor of 1000 to $z'=z\times 10^{-3}$. The values for the pulse energy $W$ and number of photons in the pump ($N_p$) and dump contributions  ($N_d$) of the SASE beam ($W=\hbar(\omega_p N_{p}+\omega_d N_d)$) are to be understood as effective values in the interaction region. Typically the beam-line losses are on the range of 80-90\%, which has to be accounted for, if comparing the quoted values to typical machine parameters (that are typically quoted without taking into account transport losses to the experimental stations). In our calculations we suppose a cylindrical-shaped focus of a cross sectional area of $s_x=7\; \mu m^2$. The number of photons on target therefore relate to the field-amplitudes of $E_{p,d}$ as $N_{p,d}=(c s_x /8\pi\hbar\omega_{p,d})\int |E_{p,d}|^2dt$. In the numerical simulations we employ SASE pulses of a rectangular envelope of $\tau_x=$100 fs duration. Both, the pump and dump pulse are considered to be overlapping in time and have the same duration. These long pulse duration ensures, that the spectral coherence (average width $\delta\omega$ of a spectral SASE spike) is smaller than the vibrational level spacing, i.e.  $\delta\omega\geq 4\ln(2)/\tau_x\approx 0.02$ eV, a prerequisite to obtain vibrationally resolved spectra (see Sec. \ref{sec:low}). The final spectrum $I_{p,d}(\omega,z)$ is obtained by Fourier transform and squaring of the corresponding field component $E_{p,d}(t,z)$.

\section{Results and discussion}\label{sec:res}
\subsection{Raman gain and amplification}

\begin{figure}
\centering
\includegraphics[width=0.4\textwidth]{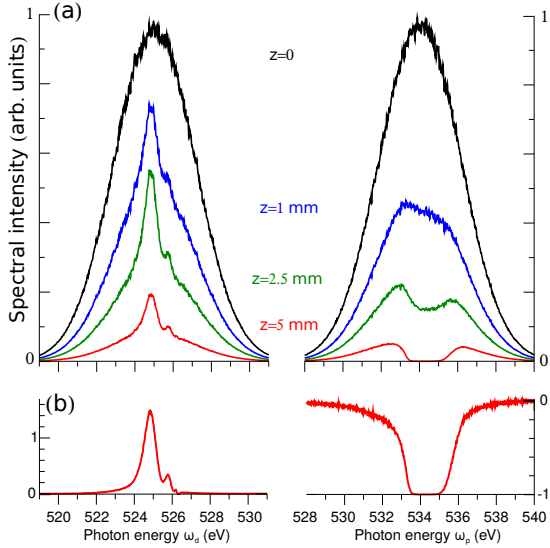}
\caption {a) Dependence of the pump $I_p(\omega,z)$ (right) and dump $I_d(\omega,z)$ (left) spectra on propagation length for the strong pump regime ($N_p(0)=10^{12}$, $N_d(0)=10^{10}$ photons per pulse, pulse duration is 100 fs). The spectral profiles at various propagation distances $z$ are averaged over 4000 individual SASE shots. The spectra are normalized to the maximum of the incoming intensities. b) Relative change of spectral intensity $[I(\omega_{i},z)e^{\sigma_{ph}Nz}/ I(\omega_{i},0)-1]$ at $z=5$ mm for the dump $\omega_d$ (left) and pump $\omega_p$ (right) field, respectively.}
\label{fig:specstrong}
\end{figure}

The first thing to investigate is the expected level of amplification of the seed pulse by SRIXS. In the atomic case \cite{Weninger2013e}, amplification of the seed photons by 6-7 orders of magnitude were observed, by pumping the pre K-edge Rydberg states. For molecular targets, generally lower Raman gain and amplification levels are expected. To quantitatively predict the Raman gain, Eqs. (\ref{weq})-(\ref{dm1}) were solved for an ensemble of 4000 pairs of SASE pump and dump pulses, for realistically achievable pulse parameters at the LCLS XFEL. We suppose $10^{12}$ photons in a 100 fs long pump pulse and $10^{10}$ photons in the dump pulse and a focal area of 7 $\mu m^2$, resulting in average intensities of 6.5$\times 10^{16}$ W/cm$^2$ and 6.5$\times 10^{14}$ W/cm$^2$, respectively.  Fig. \ref{fig:specstrong}(a) shows the ensemble-averaged spectra for the incoming ($z=0$) pair of SASE pulses (each of them is by construction a Gaussian with 5 eV width at FWHM) and the transmitted pulses for different propagation lengths through the medium. The O$1s\rightarrow \pi^*$ resonance has a relatively strong electronic dipole transition element of 0.05 a.u., that at the relatively high pump intensities of 6.5$\times 10^{16}$ W/cm$^2$ translates into a Rabi frequency of 2.6 eV and Rabi period of 1.6 fs. This implies that the pump pulse drives strong Rabis oscillations, with a period shorter than the Auger lifetime\cite{Demekhin2011a,rohringer_strongly_2012}. Propagating through the medium, the pump pulse is strongly absorbed on the O$1s\rightarrow \pi^*$ resonance in CO (broad absorption feature centred at 534.5 eV, Fig.\ref{fig:specstrong}(b)). The lifetime width of each vibrational state is $\Gamma_i=0.16$ eV, corresponding to a core-hole Auger lifetime of 4 fs. With an averaged vibrational energy spacing of 0.166 eV\cite{Coreno1999b}, the individual vibrational states should be observable in the absorption spectrum. Due to the strong nonlinear resonant coupling inducing Rabi oscillations (see Fig \ref{fig:pops}) and broadening of the absorption features, the vibrational structure is, however, washed out. The vibrational structure can however be observed in the weak pump regime (see Section \ref{sec:low}). Stimulated resonance scattering on the O$1s\rightarrow \pi^*$ transition  leads to amplification (Raman gain) on the low-energy SASE pulse (Fig. \ref{fig:specstrong}(b), left panel), and clearly a peak grows out on the Gaussian averaged spectrum at 525 and 525.6 eV photon energy (Fig.\ref{fig:specstrong}(b), left panel), as the pulses are co-propagating through the medium. Both spectral components are attenuated about one order of magnitude due to the non-resonant absorption channels (photoionization of C1s and other orbitals). Compared to the transmitted incoming SASE pulse, an amplification of 1.4 times is observable in the transmitted spectra at the peak of the SRXIS feature at 525 eV photon energy.

\begin{figure}
\centering
\includegraphics[width=0.47\textwidth]{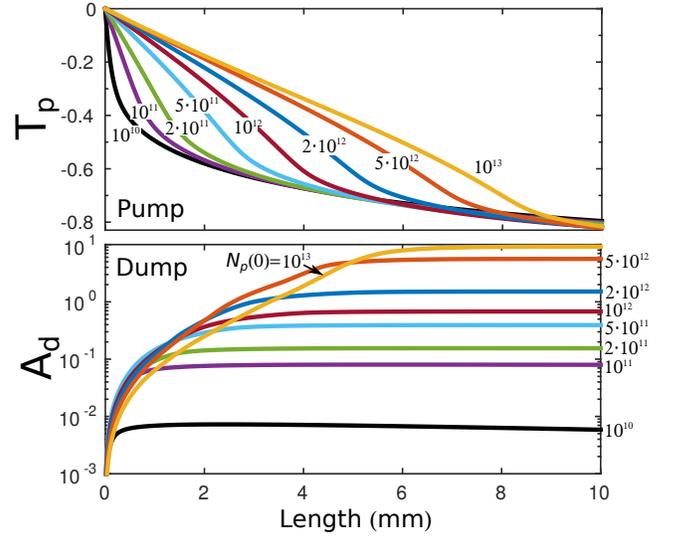}
\caption{Pulse attenuation/amplification as a function of propagation length according to Eqs. (\ref{ad},\ref{tp}). In order to clearly see the effect of the resonant scattering processes, the off-resonant absorption is subtracted from the results of the simulations (see text). Transmission and Raman amplification are shown for several incoming pump energies (see plot legends). The incoming dump pulse contains $N_d(0)=10^6$ photons. The pulse duration is 100 fs. All calculations are averaged over 128 SASE shots.}
\label{fig:gain}
\end{figure}

To estimate the average Raman gain as a function of the applied intensity, we varied the incoming number photons in the pump SASE beam. Fig. \ref{fig:gain} shows the Raman amplification of the dump pulse defined by 
\begin{equation}\label{ad}
 A_d=N_d\exp (\sigma_{ph}Nz)/N_d(0)-1,
\end{equation}
(lower panel) and the transmission of the pump pulse 
\begin{equation}\label{tp}
 T_p=N_p \exp (\sigma_{ph}Nz)/N_p(0)-1,
\end{equation}
(upper panel) as a function of the propagation distance through the medium for various numbers of incoming photons, ranging from 10$^{10}$ to 10$^{13}$. The transmission and amplification is corrected for the overall off-resonant absorption in the medium. Strong pump intensities in the nonlinear resonant coupling scheme result in high Raman gains, as it is shown in Fig. \ref{fig:gain}. Beer-Lambert's law of exponential absorption is only valid for up to $10^{10}$ photon per pump pulse. Strong resonant coupling at $N_p=10^{11}$ results in higher transmission and resonant propagation of high-intensity effective $n2\pi$-pulses through the medium\cite{eberly}. The Raman amplification initially grows exponentially, but due to the absorption of the pump radiation, exponential amplification can be maintained only up to a certain propagation distance $z$.

\begin{figure}
\centering
\includegraphics[width=0.4\textwidth]{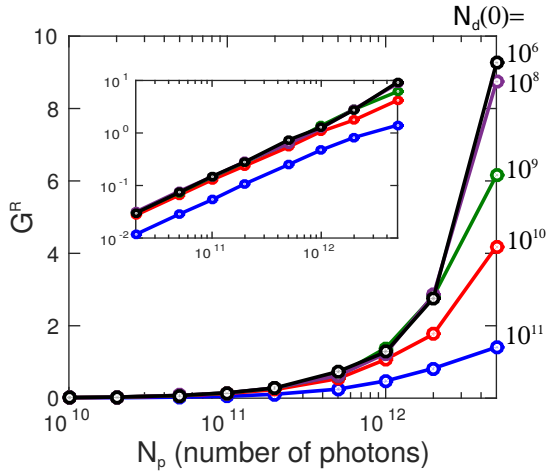}
\caption{Integrated Raman amplification $G^R$ (\ref{eq:gr}) in spectral domain 523.5-526.5 eV at $z=10$ mm as a function of the pump energy. The incoming dump pulse energies are shown in the figure legends. The off-resonant absorption was subtracted the same way as in Fig. \ref{fig:gain}. The inset presents the same dependence in log-scale on both axis, showing an exponential grows of the emission with increase of the pump energy.}
\label{fig:satcurve}
\end{figure}

To achieve high Raman gain, also the choice of the intensity of the dump (seed) pulse is important. As a measure of the Raman gain, we introduce the integrated Raman amplification $G^R(z)$ near the maximum of the emission feature at 523.5-526.5 eV (see Fig. \ref{fig:specstrong}b) as
\begin{equation}\label{eq:gr}
G^R(z)= \frac{\int_{523.5}^{526.5 }d\omega I_d(\omega,z) e^{\sigma_{ph}Nz}} {\int_{523.5}^{526.5}d\omega I_d(\omega,z=0)}-1,
\end{equation}
where $I_d$ denotes the spectral intensity of the dump pulse at position $z$ in the medium. Fig. \ref{fig:satcurve} shows the integrated Raman amplification $G^R(z)$ as a function of the number of pump photons for several values of the number of the dump (seed) photons ranging from $10^6$ to $10^{11}$. The relative amplification drops for a number of pump photons exceeding 10$^{10}$. At this particular photon number, the intensity reaches the saturation intensity on the $^1\Pi^*\rightarrow ^1\Sigma^-$ transition, i.e. the intensity is high enough so that the rate for stimulated scattering becomes comparable to the Auger-decay rate. Below this saturation regime, the amplification factor is nearly independent of the number of seed photons, corresponding to the so-called small signal gain region (see inset in Fig. \ref{fig:satcurve}). To have a realistic chance to see Raman gain in the averaged spectra, a Raman amplification of at least 50\% should be targeted. For the considered pulse durations and focal size, this implies a number of pump photons $N_p\geq 5\times 10^{11}$ and a number of dump photons $N_d\leq 10^{10}$.

\begin{figure}
\centering
\includegraphics[width=0.45\textwidth]{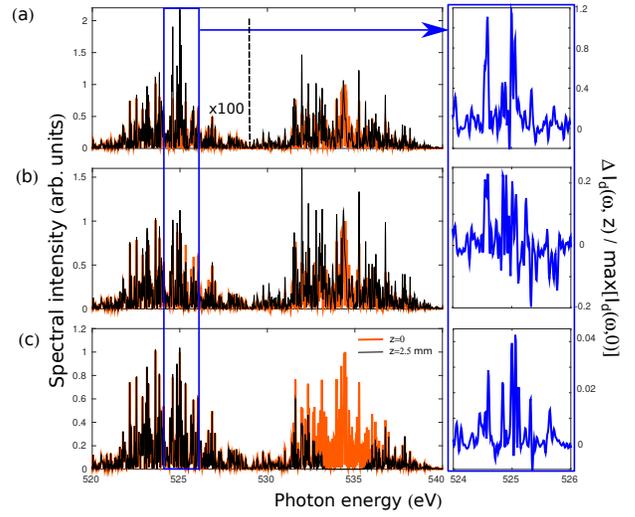}
\caption{Two-color SASE spectra at $z=0$ (thick orange lines) and $z=2.5$ mm (thin black lines) propagating distance in the CO medium; we corrected for the off-resonant absorption and all the spectra are normalized to the maximal peak intensity at $z=0$. The calculations use the same SASE spectral profile renormalized to the specific number of incoming photons: (a) $N_p=10^{12}$ photons and $N_d=10^{10}$ photons, (b) $N_p=N_d=10^{12}$, and (c) $N_p=N_d=10^{10}$ photons. The right panels show the spectral difference of the incoming and outgoing pulses $\Delta I_d(\omega,z)/\max[I_d(\omega,0)]$ normalized to the maximal peak intensity at $z=0$.}% ($\max[I_d]=6.5\times 10^{14}$ W/cm$^2$ (a,c), $\max[I_d]=6.5\times 10^{16}$ W/cm$^2$ (b)).}
\label{fig:specsase}
\end{figure}

\begin{figure*}
\centering
\includegraphics[width=0.7\textwidth]{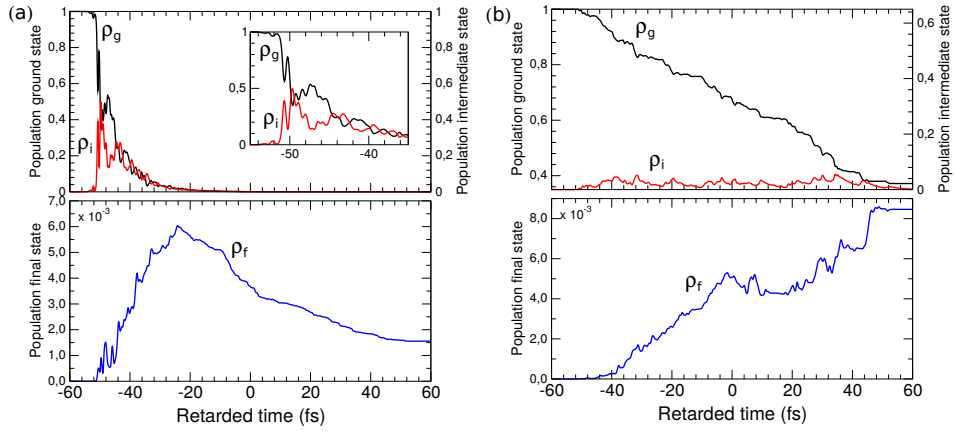}
\caption{Populations at $z=0$ for the strong pump regime $N_p=10^{12}$, $N_d=10^{10}$ (a) and weak pump regime $N_p=10^{10}$, $N_d=10^{10}$ (b). Black, red and blue lines are the total populations of the ground $\rho_g$, core-excited $\rho_i$, and final $\rho_f$ electronic states, respectively; the retarded time scale (measured relative to the center of the pump pulse) is used. The inset in plot (a) shows a zoom of the populations dynamics at the beginning of the pump pulse.}
\label{fig:pops}
\end{figure*}

The typical two-color spectrum of an incoming ($z=0$) and transmitted (for a propagation length of $z=2.5$ mm) SASE pulse can be seen in Fig. \ref{fig:specsase} for three combinations of pump- and dump-pulse energies. Intense pump fields strongly couple the ground and core-excited states and trigger nonlinear population dynamics (Rabi flopping). Fig. \ref{fig:pops} shows the occupation probabilities of the electronic ground, core-excited and final states, summed over the different vibrational states, as a function of time for a single pair of SASE pulses. Fig. \ref{fig:pops}(a) shows that in the case of strong pump intensity after $~20$ fs the ground-state is mostly depleted by core-excitation. Rabi oscillations are induced between the electronic ground and intermediate states (oscillations starting at around -48 fs). Only a small fraction (on the level of 1\%) of the core-excited population is transferred by stimulated resonance scattering to the final valence-excited state. The Raman gain is, hence, quite low and the majority of core-excited states decay by the resonant Auger effect. The strong-field limit, although favorable in terms of Raman gain, disguises the spectral information due to nonlinear resonant coupling. Rabi oscillations induce broadening of the spectral features. If one is not interested in strong-field features, and wants to obtain high-resolution spectra without altering the spectra by strong-field effects, a lower pump intensity has to be chosen. 

The case of the low pump field regime $N_p=N_d=10^{10}$ photons is presented in Fig. \ref{fig:pops}(b). The core-excited state population is rather small in this case ($\le 5$ \%). At the peaks of the pump pulse (Fig. \ref{fig:pops}(b)), the linear decrease of the ground state population is interrupted by weak Rabi oscillations between the core-excited and ground state populations. The final state's population is also small, but in the present case it shows a higher population transfer of $\sim 20$\% from the intermediate to the final state. The analysis of the population, the gain and the transmitted spectrum (Fig. \ref{fig:specstrong}) allows to determine an optimal optical density of the sample for a given set of XFEL pulse parameters. Keeping the molecular density fixed, the optimal length of the medium corresponds to roughly to 1-3 absorption lengths at the resonance. This ensures strong population of the intermediate state and maximal SRIXS amplification. With the parameters used in the simulations of Figs. \ref{fig:specstrong}, \ref{fig:pops} the optimal medium length is $\sim 3$ mm. 

\begin{figure}
\centering
\includegraphics[width=0.47\textwidth]{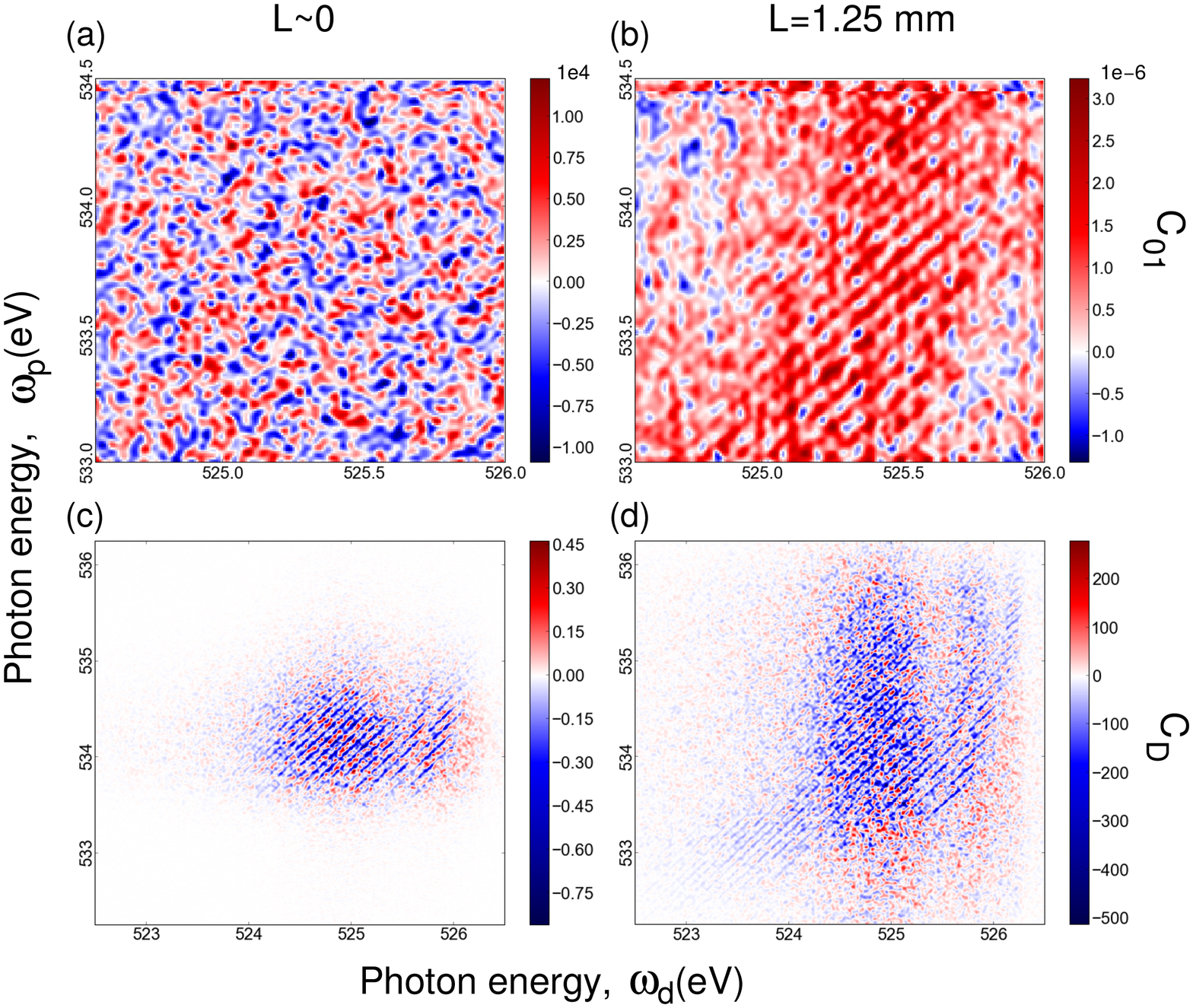}
\caption{Covariance maps of the transmitted spectra $C_{T}$ (a,b) and of the difference spectra $C_\Delta$ (c,d) for a optically thin medium (a,c) and $z=1.25$ mm (b,d) using an ensemble of 4000 spectra for the $N_p=N_d=10^{10}$ photons.}
\label{fig:covmaps}
\end{figure}

\subsection {Covariance analysis for high-resolution spectroscopy}\label{sec:low}

In the case of moderate pump-field intensities, high-resolution SRIXS photon-in photon-out spectra can be obtained by covariance analysis of the spectra \cite{Weninger2013d}. To stay in this weak-field limit, we fix the number of photons in the incoming pump pulse to 10$^{10}$, corresponding to a pulse energy of $\sim 1 \;\mu$J. Fig. \ref{fig:specsase}(c) shows the incoming and transmitted spectra for a representative pair of SASE pulses. On a linear scale the pump pulse is absorbed after a propagation distance of 2.5 mm. As evident from Fig. \ref{fig:pops}(c) the population of the intermediate state is below 5\% at the entrance of the medium, and drops down to below 0.1\% at $z=2.5$ mm (not shown here). The expected spectral Raman amplification is typically in the range below $5\%$ (see right panel of Fig. \ref{fig:specsase}(c)), and realistically unobservable in a single spectrum or the average. The SRIXS process is, however, extremely sensitive to the spiky spectral structure of the SASE pulses:  To ensure strong resonant coupling on the pump transition, the peak of a spectral spike at frequency $\omega_{p,s}$ (where $s$ counts the peaks of the stochastic spectrum) of the incoming pump field has to overlap, or be sufficiently close to a vibronic transition $\omega_{g,\{i,\nu_i\}}$ of the ground state to one core-excited intermediate state $\{i,\nu_i\}$. According to the linear dispersion of resonance scattering following from the Kramers-Heisenberg equation for second order perturbation theory \cite{gelmukhanov_resonant_1999}, a spectral spike of the dump (seed) pulse $\omega_{d,s’}$ has to energetically match the energy difference  $\omega_{d,s’}=\omega_{p,s}-\omega_{g,\{f,\nu_f\}}$, to a specific final state $\{f,\nu_f\}$. As a consequence, there is a correlation between the absorption strength of specific spikes of frequency $\omega_{p,s}$ and Raman gain of specific spikes of frequency $\omega_{d,s’}$. A covariance analysis \cite{Frasinski1989a,Zhaunerchyk2013a,Weninger2013d} unravels this correlation and is able to map-out the complete vibronic level structure of the underlying molecular system. Evidently, this information is lost by simply averaging the spectra (see Fig.\ 2).
The spectral covariance $C_{T}(\omega_p,\omega_d;z)$ at propagation distance $z$ is defined by
\begin{eqnarray}
%C_{T}(\omega_p,\omega_d;z)=&&\langle I_p^{(res)}(z,\omega_p)I_d^{(res)}(z,\omega_d)\rangle \nonumber\\
%&&-\langle I_p^{(res)}(z,\omega_p)\rangle \langle I_d^{(res)}(z,\omega_d) \rangle, \label{cov01}\\
%I_{p,d}^{\rm(res)}(z)=&&I_{p,d}(z)e^{\sigma_{ph}Nz}\;.
C_{T}(\omega_p,\omega_d;z)=&&\langle I_p(z,\omega_p)I_d(z,\omega_d)\rangle \nonumber\\
&&-\langle I_p(z,\omega_p)\rangle \langle I_d(z,\omega_d) \rangle. \label{cov01}
%I_{p,d}^{\rm(res)}(z)=&&I_{p,d}(z)e^{\sigma_{ph}Nz}\;.
\end{eqnarray}
For small Raman gain it is advantageous to define the covariance of the difference spectra between incoming and outgoing pulses:
\begin{eqnarray}
C_\Delta(\omega_p,\omega_d;z)=&&\langle \Delta I_p(z,\omega_p) \Delta I_d(z,\omega_d) \rangle \nonumber\\
&&- \langle\Delta I_p(z, \omega_p)\rangle \langle \Delta I_d(z,\omega_d) \rangle, \label{covd}\\
\Delta I_{i}(z,\omega)=&&I_{i}(z=0,\omega)e^{-\sigma_{ph}Nz}-I_{i}(z, ,\omega)\; i=p,d\nonumber
\end{eqnarray}
Experimentally, this would imply to measure both spectra $I_p$ and $I_d$ with high resolution and find a procedure to adequately renormalize the transmitted spectrum to correct for nonresonant absorption in the medium. In this analysis we, according to our absorption model in the medium, applied the normalization factor $e^{\sigma_{ph}Nz}$ to the incoming pulses. Fig. \ref{fig:covmaps} shows the covariance maps of Eq.\ (\ref{cov01}) and (\ref{covd}) for an ensemble of 4000 simulated pulses. The incoming x-ray pulses, by construction, do not show any correlations, so that the covariance $C_T(z=0,\omega_p,\omega_d)$  is that of Gaussian noise (Fig. \ref{fig:covmaps}(a)). As propagating through the medium, the correlation between absorption and emission frequencies is building up and the fine vibrational structures and linear energy dispersion of the SRXIS process are clearly visible for the covariance of the transmitted spectra for the full propagation distance of $z=1.25$ mm (Fig. \ref{fig:covmaps}(b)). The vibrational fine structure and quality of the covariance map improves considerably when looking at the covariance of the difference spectra $C_\Delta$ (Fig. \ref{fig:covmaps}(c,d)). Moreover, a clear vibrational structure is already visible for an optically thin medium (Fig. \ref{fig:covmaps}(c)).

The fine vibrational structure observed in Fig. \ref{fig:covmaps} is clearly related to the final state vibrational progression. Indeed, the spectral features of the covariance maps follows the Raman law $\omega_{p}-\omega_{d}=\omega_{g,\{f,\nu_f\}}$, where $\omega_{g,\{f,\nu_f\}}$ denotes the RIXS energy loss. The spectral resolution here is limited mainly by the average spectral width of the individual SASE spikes $\delta\omega$. %, or in other terms the spectral coherence of the SASE source.  $\delta\omega$ is inversely proportional to the duration of the SASE pulse and can be varied in an experiment.

\begin{figure}
\centering
\includegraphics[width=0.47\textwidth]{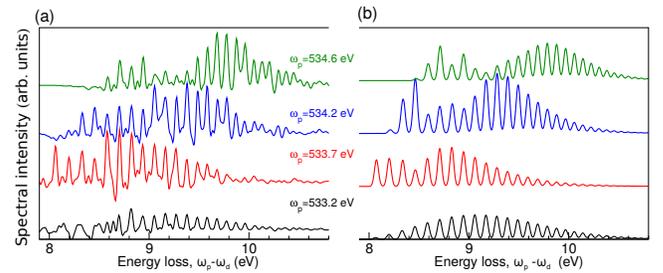}
\caption{(a) SRIXS spectra obtained with the help of the covariance map of Fig. \ref{fig:covmaps}(d) using a narrow spectral window (0.2 eV) around $\omega_{p}=532.2$, 533.7, 534.2, and 534.6 eV. (b) Conventional RIXS simulations at the narrow band (0.05 eV) excitation for the same $\omega_p$ energies as in (a).} %Energy scale is relative to the ground state energy $\omega_{gf}$.}
\label{fig:partcov}
\end{figure}

To demonstrate that the SRIXS covariance map in the weak-pulse limit recovers the vibronic level structure of the system, including excitation energies, Frank-Condon factors and transition strength, we analyze narrow spectral bands at fixed pump frequency $\omega_p$ and compare them to conventional RIXS spectra. Fig. \ref{fig:partcov} shows cuts through the covariance map $C_\Delta$ ($z=1.25$ mm) of Fig. \ref{fig:covmaps}(d) for four different excitation frequencies $\omega_p$ (a) in comparison to the energy loss spectra at those frequencies obtained by the Kramers-Heisenberg formalism (b). A good agreement between both spectral shapes of the energy loss functions is found, both in overall shape and the position of the peaks in the energy-loss spectrum. The quality of the SRIXS covariance map can be improved by increasing the size of the ensemble. Although in this proof-of-principle study, no additional information can be gained as compared to a high-resolution RIXS experiment at a third-generation synchrotron light source, the demonstration of the equivalence of both approaches in the static limit (starting from the ground state of the system) is important to validate the method. In the presently proposed form, SRIXS can be directly extended to the time-dependent pump-probe scheme, by applying the SRIXS probe process at well controlled delay times, following an optical or UV pulse, that prepared a electronic or vibrational wave packet. As in transient absorption measurements, this method could unravel coherent vibronic wave packets. 

\section{Critical assessment of feasibility -- practical issues}\label{sec:exp}

\begin{figure}
\centering
\includegraphics[width=0.47\textwidth]{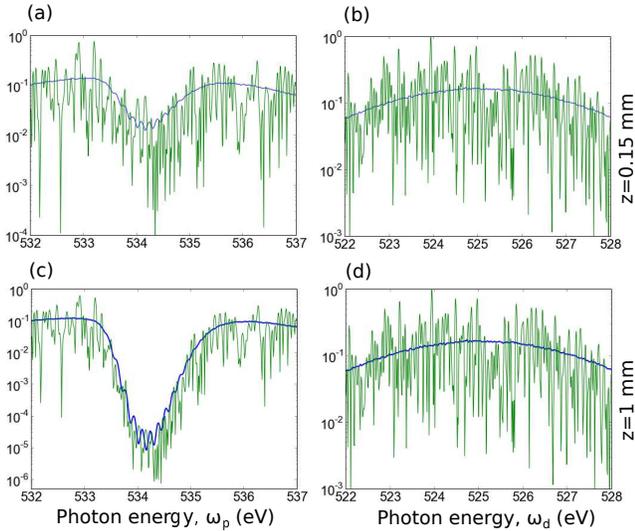}
\caption{The spectrum of a typical transmitted pair of SASE pulses (green) at the excitation (a and c) and emission (b and d) transitions and their ensemble average (blue). Panels (a) and (b) correspond to the transmitted spectrum of a propagation distance of $z=0.15$ mm and (c) and (d) $z=1$ mm. Due to strong resonant absorption of the pump radiation a larger dynamic range of the spectrometer set-up is required for the longer medium.}
\label{fig:dynrange}
\end{figure}

\begin{figure*}
\centering
\includegraphics[width=0.9\textwidth]{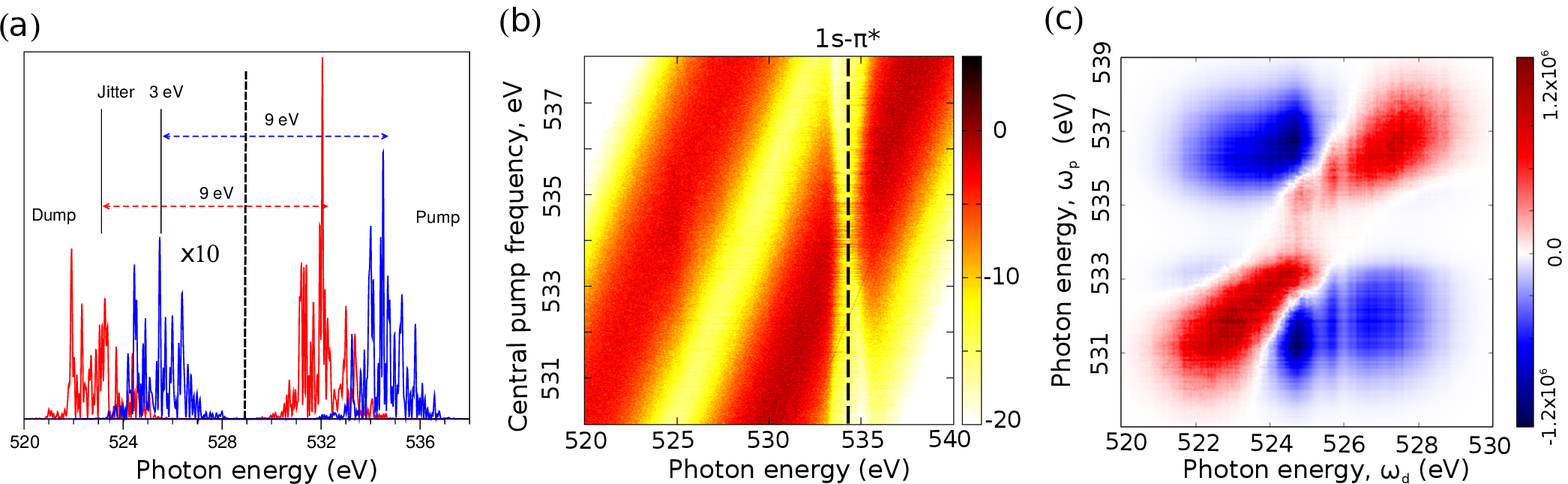}
\caption{Effect of photon-energy jitter on SRIXS measurements. (a) Example of a pair of two-color incoming SASE pulses $I(\omega,0)$ with a relative energy difference (jitter) of 3 eV. (b) 2000 simulated SRIXS spectra in logarithmic scale normalized by the averaged pump intensity $\log(I(\omega)/\overline I_p$, $\overline I_p=6.5\times 10^{16}$ W/cm$^2$ as a function of central photon energy of the pump component in logarithmic scale, that was varied by $\pm 4$ eV. (c) Covariance map $C_{T}$ of Eq. (\ref{cov01}) accounting for the pulse energy jitter. The vibrational fine structure at $\omega_d = 525-526$ eV and $\omega_p= 533-534$ eV (see Fig.\ref{fig:covmaps}(c)) is lost due to the neighboring strong features reflecting pulse energy jitter. The pulse parameters are adapted from the recent experiment \cite{lcls14expTBP}: $N_p=10^{12}$, $N_d=10^{11}$ photons, pulse duration is 100 fs, $z=5$ mm.}
\label{fig:jitter}
\end{figure*}

The theoretical predictions for the SRIXS in molecular gas-phase targets discussed in the previous section are based on numerical simulations using realistic experimental conditions, that are in principle available at present-day XFEL facilities, in particular the LCLS \cite{Lutman2013}. The European XFEL facility, currently under construction and planned to start operation in 2017, as well as the planned LCLS II project would even offer higher peak fluxes and, hence widens the applicability of our proposed method. Our numerical estimates to describe the molecular medium are based on accurate {\it ab initio} calculations for the transition dipole moments and experimental potential energy curves and the SASE x-ray radiation is simulated with the help of stochastic model giving reasonable agreement with the recorded SASE spectra. Our predictions can therefore be considered as quantitatively predictive and realistic, and the observation of SRIXS in CO has been recently experimentally attempted \cite{lcls14expTBP}, so far, albeit, unsuccessfully due to unfavorable conditions of the total pulse energy on sample, that could be achieved in this attempt. Here, we want to address a few practical issues and limitations, to give a fair assessment of a possible future realization of SRIXS. We will discuss restrictions and typical performance of the XFEL sources themselves (pulse energy jitter, pulse durations, photon-energy jitter, etc.), as well as typical beam-line parameters, such as the focus size, mirror losses, resolution and dynamic range of the spectrometer set-up, etc. 

Although theoretical covariance maps can reproduce the RIXS high-resolution spectra for low pump fluence, realistically, one needs a Raman gain in the range of at least 0.1 to establish experimental covariance maps of decent quality. Thus, at least $10^{12}$ photons on target (corresponding to 100 $\mu$J pulse energy on target for a photon energy of 534 eV) in a focal spot of 1.5 $\mu$m radius are required for SASE pulses of 100 fs duration (see Fig. \ref{fig:gain}). These parameters can be realistically achieved at the LCLS facility.

In addition to these quite demanding parameters for the output of the FEL radiation, the conditions on the spectroscopy set-up is quite stringent. Fig. \ref{fig:covmaps} shows theoretical covariance maps, calculated from ideal transmitted spectra, i.e. no shot-noise in the photon detection, background, finite dynamic range of the spectrometer setup, etc. was taken into consideration. To experimentally obtain high-quality covariance maps one needs a spectrometer set-up that\\
\begin{itemize}
\item{(i) has high spectral resolution, on the scale of the spectral coherence of the XFEL source}
\item{(ii) has high dynamic range (at least 4 orders of magnitude), so that the weak signal of the absorbed spectral regions, and spectral spikey structure of the SASE pulses can be resolved on a logarithmic scale, along with the amplified spectral regions}
\item{(iii) high read-out rates, to ensure that a large ensemble of single-shot spectra can  be recorded in a reasonable amount of time.}
\end{itemize}
ad (i): Today, only few beamlines are equipped with spectrometers of resolving power $E/\Delta E\geq 10000$ in soft x-ray range \cite{ghiringhelli} and at most XFEL beam lines, a high-resolution spectrometer is not part of the instrumentation. With certain transportable spectrometers a resolving power of  $E/\Delta E\sim 5000$ can be achieved, albeit at signal loss, and loss of the dynamic range, since these resolutions require operation of the grating-based spectrometer at higher diffraction orders.\\
(ii) In order to get high quality SRIXS maps one has to record the spiky structure of each individual XFEL pulse with high accuracy. Low-noise, back-illuminated x-ray CCD cameras have a dynamic range of typically $\sim 10^4$. For strong absorption resonances (as it is the case for the present study on CO) a higher dynamic range would be required. Fig. \ref{fig:dynrange} shows simulated spectra of a pair of transmitted SASE pulses at two different length settings of the medium. For the thin medium (Fig.\ref{fig:dynrange} (a, b)), that also shows smaller Raman gain, a dynamic range of 10$^4$ would be required. For the medium with higher optical density (Fig.\ref{fig:dynrange} (c, d)), a dynamic range of 10$^6$ is required, to quantitatively monitor the absorption on the strong resonance. Cutting-off the spectrum at a minimum threshold intensity  (simulating the finite range of the detector) leads to a smearing the covariance map and, hence, loss of spectral information.\\ 
(iii) In order to improve the quality of the covariance analysis a large number of the individual shots has to be acquired in a reasonable amount of time. The LCLS XFEL has a maximum repetition rate of 120 Hz and the upcoming European XFEL will have a repetition rate greater than 2 kHz.  The read-out rates of commercially available small pixel ($\sim 10 \times 10 \mu$m$^2$) x-ray CCD cameras are however only a few Hz, which is the main limiting factor for collecting a better statistics. Improving the dynamic range,  read-out rate and maintaining a small pixel size will be future technological challenges for high-resolution spectroscopy at XFEL sources.

All our calculations presented in the last section are for ideal SASE pulses, simulated by Gaussian noise. The pulse energies of the individual pulses were renormalized to a constant value. In reality, SASE pulses from XFELs have additional shot-to-shot changes. Pulse-energy fluctuations can be monitored on a shot-to-shot basis and could be filtered out by partial covariance analysis. A more critical issue is the jitter (shot-to-shot variation of the electron-bunch energy), that results in a jitter of the central photon energy of the pulses. Typically the shot-to-shot energy jitter is on the order of 1$\%$. For the two-pulse SASE operation, the energy difference of the two pulses is nearly constant from shot-to-shot, but the electron-bunch energy jitter in our last experiment \cite{lcls14expTBP} resulted in stochastic shifts of the pair of pulses in a range of $\pm$4 eV (see Fig.\ref{fig:jitter}(a) showing two randomly generated two-color SASE pulses with the energy jitter about 3 eV). To illustrate the influence of the photon-energy jitter on the covariance technique, we simulated 2000 transmitted spectra, assuming a photon-energy jitter in the range of $\pm$ 4 eV for pulse energy $N_p=10^{12}, \; N_d=10^{11}$ photons, similar to the pulse parameters obtained in the experiment. Fig.\ref{fig:jitter}(b), shows the log-scale spectra as a function of the central photon energy of the pump component ranging from 530 to 538 eV (in an actual experiment, this quantity is determined on a shot-to-shot basis by monitoring the electron-bunch energy). The strong, broad absorption $1s\rightarrow\pi^*$ resonance is clearly visible at 534 eV.  Fig.\ref{fig:jitter}(c) shows the corresponding covariance map $C_T$ of the same data set. The energy correlation of the pump and dump frequency is clearly seen as the diagonal feature, showing a width of $~ 3$ eV, that corresponds to the width of a single SASE component.  The position of the $1s\rightarrow\pi^*$ absorption resonance, is visible as correlated region at the crossing point 534/525 eV, for which the covariance drops. The spectral region between the pump and dump pulse become visible as the large-scale structures of negative covariance, showing a scale of 5-6 eV, that corresponds to the dark spectral region in between the pair of SASE pulses. Fine modulations of the large-scale structures are due to unconverted statistics. Comparing to the ideal covariance maps of Fig.\ref{fig:covmaps}, the Raman signal and the vibrational fine structure (diagonal fine stripes the covariance map) is completely buried in the dominating covariance features of the jittering SASE pulses. A comparison of the absolute scale of the covariance shows that the covariances due to the Raman scattering process are on the order of 10$^{-6}$ (see Fig.\ref{fig:covmaps}), the overall scale of the covariance in Fig.\ref{fig:jitter}(c) is however of order 10$^6$, which already demonstrates, that simple covariance analysis is not sufficient. A way to circumvent this problem would be to record a large quantity of single-shot spectra and plot partial covariances\cite{Frasinski1989a, Zhaunerchyk2013a} with respect to the central-photon energy (electron-bunch energy of the accelerator). 

\section{Conclusions}\label{sec:conclusions}

We presented a detailed feasibility study on stimulated resonant inelastic x-ray scattering (SRIXS),  also referred to as stimulated resonant x-ray Raman scattering with self-amplified spontaneous emission (SASE) pulses from x-ray free-electron lasers. Our theoretical model is quantitatively predictive and includes the treatment of the electronic and vibrational degrees of freedom of the molecular system, by restricting the molecular density matrix to a number of resonantly coupled electronic states. The equations of motion for the density matrix are self-consistently coupled to the Maxwell Equations in the one-dimensional paraxial approximation, so that propagation effects, absorption and stimulated resonant scattering can be treated. In this way, quantitative predictive spectra of the transmitted x-ray pulses are determined, that are the basis for discussing possible experiments at XFEL facilities that are available presently, or in the near future.  A detailed discussion of SRIXS in an ensemble of CO molecules at the O$1s\rightarrow\pi^*$ core-excited resonance is presented. The pump intensity and optical density of the gas medium can be optimized to maintain a high Raman-gain regime throughout the whole medium. For CO at density-length products of $Nz\approx 2.5\times 10^{-18}$ cm$^{-2}$, an intensity on target of $\sim 6.5\cdot 10^{16}$ W/cm$^2$ (corresponding to $10^{12}$ photons on a 7 $\mu m^2$ cross sectional area within a pulse of 100 fs duration) is required to achieve a sizeable occupation of the core-excited state and inducing high Raman gains. In this regime, Raman gains of roughly an order of magnitude in the spectral intensity can be obtained, however, in the strong nonlinear coupling region of the ground state to core-excited state transition. This results in broadened spectral features that undermine high resolution spectroscopy as a diagnostic method, but open up the possibility to study nonlinear optical effects, such as Rabi oscillations and saturated stimulated electronic x-ray Raman scattering, in the x-ray region. More sophisticated statistical analysis techniques, as applied in nonlinear stochastic spectroscopy with optical light \cite{doi:10.1021/jp310477y}, might be a pathway to extract nonlinear x-ray response functions, but need to be developed.

In the weak-field limit, SRIXS can serve as high-resolution spectral probe process of the underlying molecule, similar to conventional RIXS at third generation x-ray light sources. With SASE pulses from XFELs, high-resolution SRIXS photon-in photon-out spectra can be obtained by the acquisition of an ensemble of single-shot stochastic spectra and subsequent covariance analysis of the spectra. The energy resolution of this method is determined by the spectral coherence of the incoming SASE pulses (the average spectral spike width that is proportional to the inverse pulse SASE pulse duration), so that with appropriately long pulses (pulse duration $>$ 50 fs) vibrational resolution can be achieved.

SRIXS has several advantages over conventional RIXS. Tuning the dump (seed) pulse can be tuned to a particular electronic transition,  so that otherwise 
“dark” or weak electronic transitions can be enhanced and accurately measured, that are otherwise buried in the RIXS signal of the strongest transitions. In contrast to conventional time-resolved RIXS at FELs, SRIXS does not rely on the availability of a monochromator. By collecting an ensemble of SRIXS spectra of $~$5000 pairs of broad band SASE pulses, a complete photon-in photon-out spectrum (SRIXS map) can be achieved. The signal is detected in forward direction (homodyne detection with the incoming fields) so that high-resolution single-shot spectra can be recorded for also optically thin samples. A critical assessment of the feasibility of the experiment and the required experimental parameters was presented.

The proposed technique can be extended to larger molecular systems and to the hard x-ray spectral range, opening new doors for studying ultra fast nuclear dynamics using the advantage of high element and site selectivity of x-rays. Moreover, our theoretical study show that stimulated electronic x-ray Raman scattering, the building block of several nonlinear x-ray pump-probe spectroscopic techniques \cite{PhysRevLett.89.043001,doi:10.1021/jz501966h} should be accessible with currently achievable x-ray intensities. These methods require, in principle, accurately timed, transform limited x-ray pulses of attosecond duration. Several theoretical proposals to create attosecond FEL pulses exist \cite{PhysRevSTAB.8.050704,PhysRevSTAB.9.050702, PhysRevLett.92.224801, PhysRevLett.114.244801, PhysRevLett.113.024801}, but so far remained experimentally unexplored.  With the advancement of XFEL sources in seeding \cite{PhysRevLett.114.054801,PhysRevLett.111.114801,Deninno}, and seeded two-pulse schemes \cite{allaria, PhysRevLett.113.254801}, two-pulse attosecond FEL operation might, however, be available in the near future. 

\section*{Acknowledgements}
We acknowledge financial support from the Max Planck Society; VK also acknowledges financial support from the Knut and Alice Wallenberg Foundation (Grant No. KAW-2013.0020) for the project “Strong Field Physics and New States of Matter”. The simulations were performed on resources provided by the Swedish National Infrastructure for Computing (SNIC 2015/1-69).

%\bibliography{rsc-02} %your .bib file


\end{document}